\documentclass[twocolumn]{aastex701}
\journalinfo{Accepted for publication in The Astrophysical Journal Letters}

\graphicspath{{figs/}}
\usepackage{bm}

\begin{document}

\title{Existence of Distinct Strong and Weak Shear Regimes of \\
Self-Generated Turbulent Magnetic Reconnection}

\author[0009-0005-3865-0281]{Qihui Ming}
\affiliation{School of Mathematics and Statistics, University of St Andrews, St Andrews, KY16 9SS, U.K.}
\email{qm7@st-andrews.ac.uk}

\author[0000-0001-5690-2351]{Alexander J. B. Russell} 
\affiliation{School of Mathematics and Statistics, University of St Andrews, St Andrews, KY16 9SS, U.K.}
\email{ar51@st-andrews.ac.uk}

%%%%%%%%%%%%%%%%%%%%%%%%%%%%%%%%%%%%%%%%%%%%%%%%%%%%%%%%%%%%%%%%

\begin{abstract}
Magnetic reconnection can become fast in three-dimensional turbulent layers. However, the parameters and mechanisms that control the fast state are not yet fully clear. Existing guide-field simulations of self-generated turbulent reconnection show locally coherent flux-rope structures embedded in a stochastic layer, motivating a flux-rope-mediated picture, with reconnection rates of order $V_{\rm rec}/V_A \sim 0.01$ in MHD. Here, we test whether and how magnetic shear changes the reconnection regime, using 3D resistive-MHD simulations at $S_L=10^5$. We vary the guide field while keeping the initial reconnecting field strength and rate normalization fixed, comparing a guide-field, weaker-shear configuration with the anti-parallel, strong-shear limit. We find that the strong-shear regime is qualitatively distinct from the flux-rope-mediated guide-field regime. Rather than retaining a narrow layer and locally coherent flux-rope structures, the current layer develops short-wavelength corrugations that broaden it substantially, and the reconnection layer electric field is sustained by fluctuation EMF. The reconnection rate increases to $V_{\rm rec}/V_A \sim 0.03$, about three times the guide-field value. These results indicate that magnetic shear can select the mechanism controlling fast 3D turbulent reconnection, shifting the system from a guide-field regime consistent with flux-rope mediation toward a very fast turbulence-broadened regime. This transition provides a novel explanation for the sudden switch-on of fast reconnection in eruptive solar flares.
\end{abstract}

\keywords{\uat{Solar magnetic reconnection}{1504} --- \uat{Solar flares}{1496} --- \uat{Magnetohydrodynamical simulations}{1966} --- \uat{Magnetohydrodynamics}{1964}}

%%%%%%%%%%%%%%%%%%%%%%%%%%%%%%%%%%%%%%%%%%%%%%%%%%%%%%%%%%%%%%%%%%%%%%%%%%%%%%%%

\section{Introduction} 

Magnetic reconnection is a key energy release process in many astrophysical phenomena, including solar flares. The classic Sweet--Parker reconnection model \citep{parker1957, sweet1958} agrees with current sheet geometry in eruptive solar flares \citep[e.g.][]{warren2018,cheng2018} but the reconnection rate it predicts is far too slow to explain the rapid energy release in high-Lundquist-number plasmas. This has led to extensive interest in fast reconnection and its triggering, with it now being well established that 2.5D MHD systems with high-Lundquist numbers achieve reconnection rates of order $V_{rec}/V_A\sim0.01$ through the plasmoid-mediated mechanism \citep{shibata2001,bhattacharjee2009,cassak2009,huang2010,uzdensky2010}, while the Hall term allows collisionless systems to reach larger reconnection rates of order $\sim0.1$ \citep[][and references therein]{birn2001}.

In 3D, another route to fast reconnection is turbulence, which can be self-generated within the reconnection layer. \citet{lazarian1999} proposed that turbulence causes field-line wandering, which broadens the outflow independently of resistivity, and hence enables fast reconnection. Many 3D simulations have subsequently confirmed occurrence of fast reconnection rates in MHD, both by artificially-driven turbulence \citep{kowal2009, vicentin2025} and self-generated turbulence \citep{oishi2015,huang2016,beresnyak2017,beg2022}, with the MHD simulations that spontaneously generated turbulence displaying reconnection rates of order $\sim 0.01$. Particle-in-cell and Hall MHD simulations also show the spontaneous generation of 3D turbulence inside reconnection layers, with the collisionless reconnection rates exceeding those for MHD \citep{daughton2011,daughton2014,huang2024}.

Many turbulent reconnection simulations develop flux ropes inside the reconnection layer \citep{daughton2011, huang2016, beg2022}. Those field-line bundles can remain coherent locally despite being turbulently dispersed over longer distances  \citep{russell2025} and they are often viewed as the 3D analogue of plasmoids. These observations complicate a purely stochastic-broadening picture, because locally coherent structures can regulate the reconnection rate, resulting in flux-rope-mediated turbulent reconnection \citep{russell2025}, which extends the principles of 2D plasmoid-mediated reconnection to 3D systems with turbulence.

Unresolved issues for magnetic reconnection with self-generated turbulence include mapping its different regimes and developing a complete understanding of what controls them. These goals are similar in spirit to the reconnection phase diagram for 2D systems \citep{ji2011}. While progress is being made for 3D systems via parameter studies \citep[e.g.][]{leake2020,daldorff2022,leake2024,liang2025}, the full set of parameters governing self-generated turbulent reconnection rates in 3D may be more extensive than those controlling 2D reconnection, and it is currently unknown what parameters are most important.

An especially interesting parameter is the magnetic shear across the reconnection layer (equivalently the guide field strength). From a theoretical standpoint, \citet{russell2025} argued that magnetic helicity supports the formation of locally coherent flux ropes in flux-rope-mediated reconnection and speculated that a different reconnection regime may exist in the anti-parallel limit, which has zero magnetic helicity. This proposal contrasts with 2.5D reconnection, for which the guide field does not change the reconnection rate \citep[e.g.][]{ji2011}. Our key question is therefore whether increasing the magnetic shear to the anti-parallel limit changes the mechanism that controls self-generated turbulent reconnection and thus its global properties.

In eruptive solar flares, the magnetic shear across the post-CME current sheet increases over time, as field lines are stretched by the eruption \citep{dahlin2022}. Thus, a transition between reconnection regimes, controlled by the increasing magnetic shear, would be a compelling explanation for why the flare impulsive phase and most rapid acceleration of the CME commence several minutes into the eruption.

Here we use a pair of 3D resistive-MHD simulations, both having the same reconnecting field strength, to compare a guide-field, weaker-shear configuration with the anti-parallel, strong-shear limit. We examine the reconnection rate, layer structure, and the electric field contributions sustaining the reconnection, and find that magnetic shear indeed changes the mechanism that controls fast self-generated turbulent reconnection. The reconnection rate increases by a factor greater than three when the reconnecting magnetic field is anti-parallel, offering a novel explanation for the switch-on of the fastest magnetic reconnection and impulsive phenomena during solar eruptions.

\section{Simulation Setup} \label{sec:setup}

We perform 3D resistive-MHD simulations with the \texttt{Lare3d} code \citep{arber2001} at a Lundquist number $S_L = V_A L /\eta = 10^5$. The initial condition consists of two large-scale flux tubes that reconnect at the $y=0$ midplane, following \citet{huang2016,huang2024}, \citet{beg2022} and \citet{vicentin2025}. The full simulation domain is $[-0.5,0.5]^3$, but figures will focus on the narrow reconnection layer between the two flux ropes, $|y| \leq 0.05$. Variables are normalized such that the initial density, strength of the reconnecting magnetic field component and upstream $V_A=B_x/\sqrt{\mu_0 \rho}$ are equal to one, hence the reported reconnection rates are equivalent to $V_{\rm rec}/V_A$. The initial velocity is zero apart from a low-amplitude random perturbation used only as a seed, with no imposed inflow nor turbulent forcing. In the current study, the initial $B_z$ is a uniform guide field and force balance is achieved via the thermal pressure. Between the two simulations, we vary only the strength of the uniform guide-field component, keeping the initial reconnecting field strength, pressure distribution, and rate normalization fixed. This moves the system between a guide-field, weaker-shear configuration and the anti-parallel, strong-shear limit. 

As a guide-field control, we use $B_z = 0.75$, matching \citet{beg2022}, but resolve the current layer with a stretched grid that maintains constant $\Delta y \approx 10^{-4}$ throughout $|y| \leq 0.05$. This run reproduces the self-generated turbulent reconnection (SGTR) phenomenology analyzed by \citet{beg2022}. However, at the increased resolution, we obtain a quasi-steady reconnection rate near $\sim 0.01 V_A$, agreeing with the simulation results of \citet{huang2016}, who also used $\Delta y \approx 10^{-4}$ at the midplane, and the theoretical flux-rope mediated value predicted by \citet{russell2025}. We therefore use this well-resolved guide-field run as the baseline for comparison with the anti-parallel run.

In the anti-parallel run, $B_z = 0$. This value implies that the large-scale flux tubes are kink unstable (the Kruskal–Shafranov factor is zero). Fortunately, however, this global mode grows relatively slowly and does not disturb the reconnection layer until $t \gtrsim 5$. This allows us to use the interval $t<5$ to study more rapid reconnection-layer processes that generate and sustain fast turbulent reconnection before the slower global evolution disturbs the layer.

\section{Results} \label{sec:results}

The two cases initially follow a similar 2.5D evolution, but separate as the initial translational invariance is broken. At later times, they produce turbulent states with different reconnection rates, reconnection layer structures, and electric-field balances.

\begin{figure*}[htpb]
    \centering
    \includegraphics[width=0.85\textwidth]{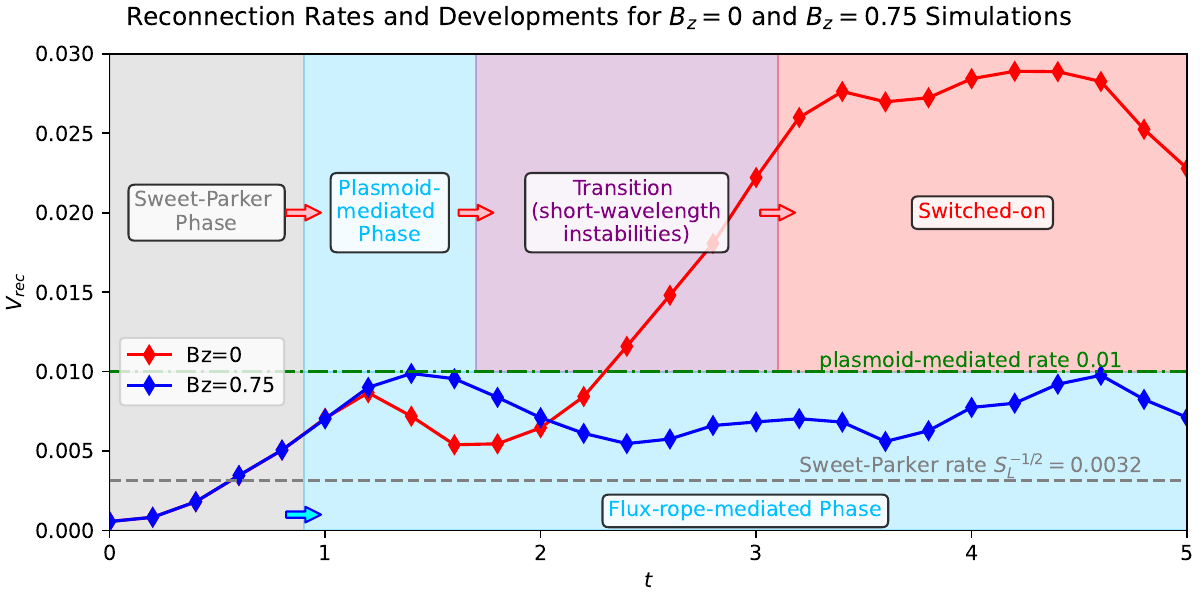}
    \caption{\textbf{Switch-on of turbulent reconnection rates.} Temporal evolution of the reconnection rate $V_{\text{rec}}$ for the anti-parallel (red) and guide-field (blue) cases. 
    Both runs initially rise from a Sweet--Parker baseline toward $V_{\rm rec}/V_A \sim 0.01$. 
    The guide-field run remains near this level, whereas the anti-parallel run undergoes a further increase and reaches a sustained rate about $V_{\rm rec}/V_A \sim 0.03$.
    }
    \label{fig:rate}
\end{figure*}

Figure~\ref{fig:rate} shows the evolution of the reconnection rate, measured from the reconnected flux using the same method as \citet{huang2016} and \citet{beg2022}.
Both runs begin at a slow Sweet--Parker rate, which increases towards the canonical plasmoid-mediated reconnection rate of $V_{\rm rec}/V_A \sim 0.01$. The simulations then depart from each other around $t=1.2$ as the initial 2.5D symmetry breaks and turbulence develops.
The reconnection rate for the guide-field run remains near 0.01 for the rest of the simulation, consistent with previous simulations \citep{huang2016,beg2022} and theory \citep{russell2025}. 
The reconnection rate for the anti-parallel run, however, undergoes a further increase beginning at $t\sim2$, and reaches a sustained rate of $V_{\rm rec}/V_A \sim 0.03$ during the quasi-steady interval from $t=3.4$ to $4.6$. 

Additional simulations with intermediate guide field values have confirmed that the two cases plotted in Figure~\ref{fig:rate} are representative. The faster reconnection regime (for weak guide fields) persists with initial guide fields up to at least $B_z=0.05$, showing that the new regime we report does not rely on an exactly antiparallel magnetic field. When the guide field is weak but non-zero, the fastest reconnection commences earlier and more sharply than it does in the $B_z=0$ case. Preliminary work completed so far indicates that the transition between reconnection regimes is reasonably sharp with respect to the guide field. We intend to examine the transition between two regimes in greater depth in a separate paper.

\begin{figure*}[htpb]
    \centering
    \includegraphics[width=0.48\textwidth]{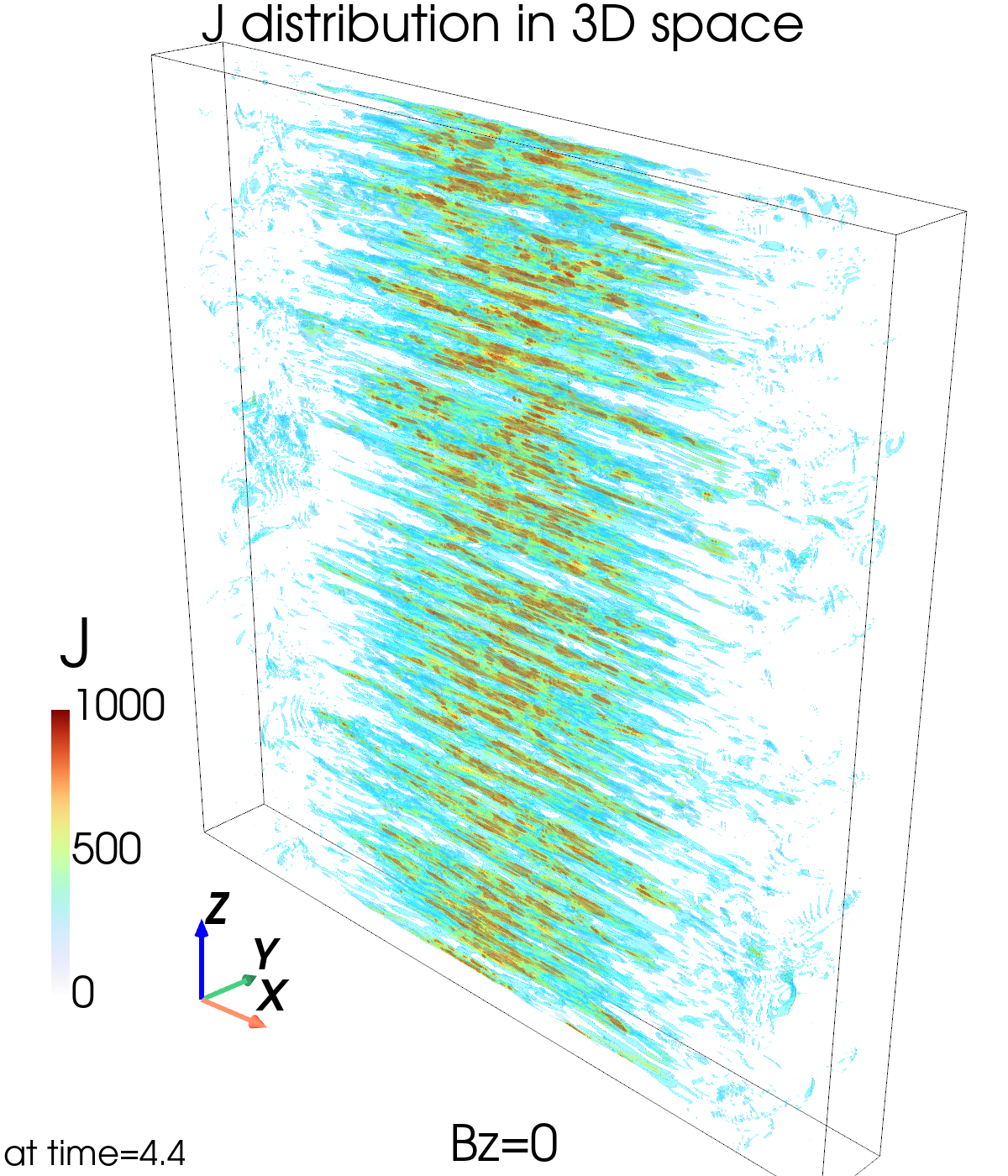}
    \hfill
    \includegraphics[width=0.48\textwidth]{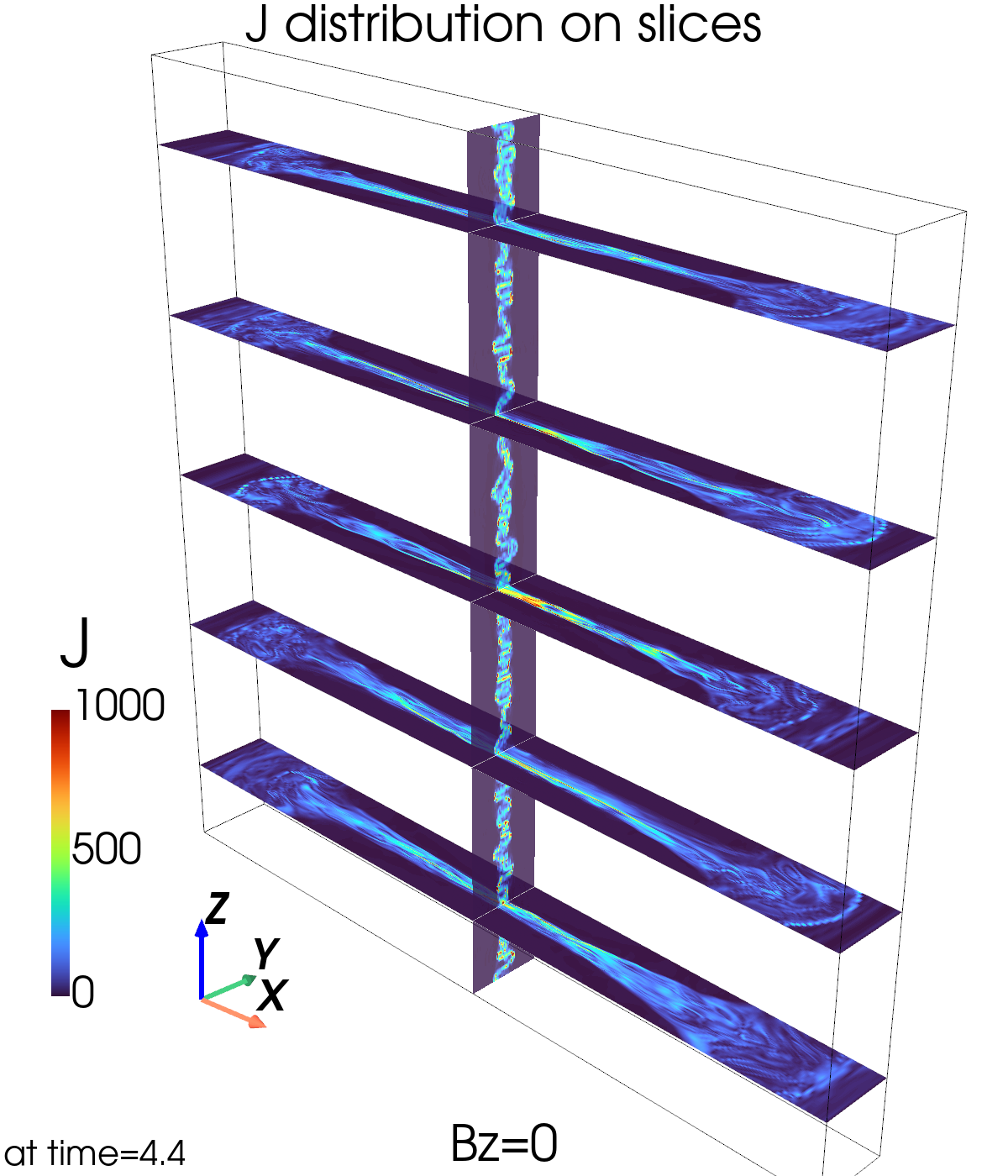}
    
    \vspace{0.5em}
    \includegraphics[width=0.48\textwidth]{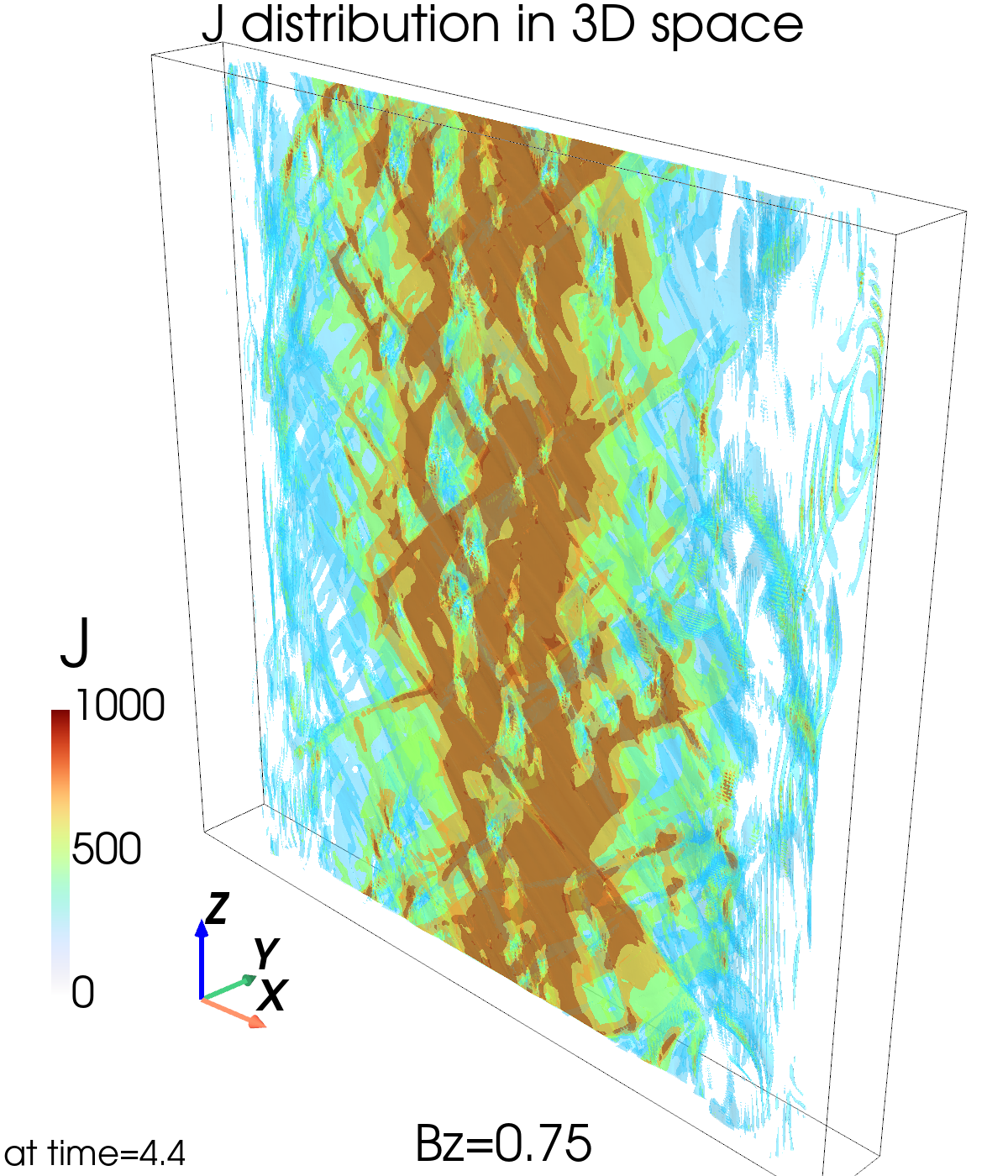}
    \hfill
    \includegraphics[width=0.48\textwidth]{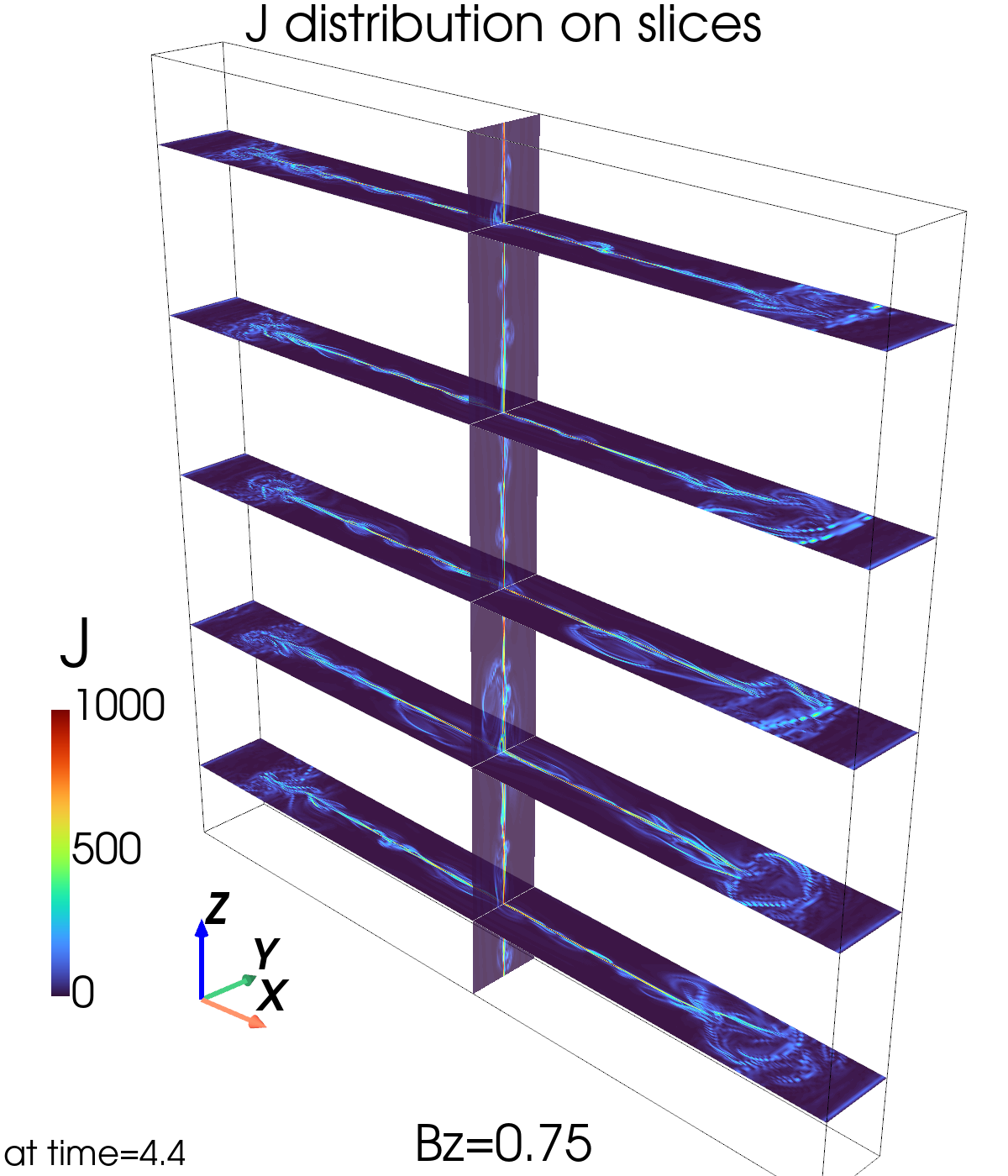}

    \caption{Distribution of current density magnitude $|\bm{J}|$, shown in 3D (left panels) and on representative slices (right panels), for the anti-parallel (top panels) and guide-field (bottom panels) simulations. 
    The displayed region is the inner subdomain $[-0.5,0.5]\times[-0.05,0.05]\times[-0.5,0.5]$ containing the reconnection layer. 
    In the anti-parallel case, the layer is broader and filled with current density, due to the development of short-wavelength corrugations seen in the vertical slice. 
    In the guide-field case, the strongest current structures remain narrower and more organized, consistent with the morphology reported by \citet{beg2022}.
    Interactive 3D versions are available online at doi: \href{https://doi.org/10.57967/hf/11330}{10.57967/hf/11330}.}
    \label{fig:j}
\end{figure*}

The structures of the reconnection layers for the two simulations are qualitatively different, with Figure~\ref{fig:j} showing clear structural differences in the distributions of current density.
In the guide-field case, current isosurfaces show locally coherent flux-rope structures, which align with the magnetic field on both sides of the current layer, while slices through the domain reveal thin and intense current sheets. This structure is consistent with previous simulations that used significant $B_z$ \citep[e.g.][]{daughton2011,huang2016,beg2022}. As has been described previously, in these cases, the dynamic flux ropes broaden the effective reconnection outflow compared to the Sweet-Parker prediction, such that the effective inverse aspect ratio of the global reconnection layer matches the reconnection rate of $\sim0.01$, while at smaller scales, the current sheet fragments locally reconnect magnetic fields at the same rate \citep[][and references therein]{russell2025}. 

The anti-parallel case, by contrast, develops highly fragmented current isosurfaces, and the vertical slice through the reconnection layer shows short-wavelength corrugations of the current sheet that substantially broaden the reconnection layer -- to several times the effective thickness obtained for the guide field case. The horizontal slices further illustrate the qualitative differences, with the anti-parallel case developing a broad reconnection layer with current distributed across it, contrasting with the thinner current sheet of the guide field case.

To identify what sustains the enhanced reconnection electric field, we decompose the mean Ohm's law,
\[
\langle \bm{E} \rangle
=
-\langle \bm{v}\rangle \times \langle \bm{B}\rangle
-\langle \tilde{\bm{v}} \times \tilde{\bm{B}}\rangle
+\eta\, \langle \bm{J}\rangle,
\]
where $-\langle \bm{v}\rangle \times \langle \bm{B}\rangle$ is the mean-flow induction and $-\langle \tilde{\bm{v}} \times \tilde{\bm{B}}\rangle$ is the fluctuation electromotive force (EMF). 
The resulting mean $\langle E_z \rangle$ provides a useful proxy for the reconnection electric field \citep{huang2016,beg2022} and is consistent with the reconnection rate independently measured from the reconnected magnetic flux. 
Near to $x=0$, the mean-flow term transports magnetic flux into the layer, while the fluctuation EMF and resistive term sustain $\langle E_z \rangle$ inside the reconnection layer. We remark that $-\langle \tilde{\bm{v}} \times \tilde{\bm{B}}\rangle$ is often referred to in the literature as the turbulent EMF; however, it arises from correlations between dynamic fluctuations around the mean state without necessarily requiring that the fluctuations are turbulent. 

Means were evaluated over $z$, as the simulations are statistically invariant in this direction, with additional averaging over $-0.1\leq x \leq 0.1$ to improve signal to noise. Spatial averaging \citep[e.g.][]{che2011,huang2016,beg2022} is appropriate when the reconnection layer does not develop additional longer wavelength perturbations in this direction \citep{le2018}, which holds true for these simulations. Time or ensemble averages should in principle produce similar results; however, spatial averaging is better justified for these simulations, since the experiments evolve gradually over time as the large scale flux ropes reconnect, and producing large ensembles of 3D simulations is currently impractical due to cost.

\begin{figure}[htpb]
    \centering
    \includegraphics[width=\columnwidth]{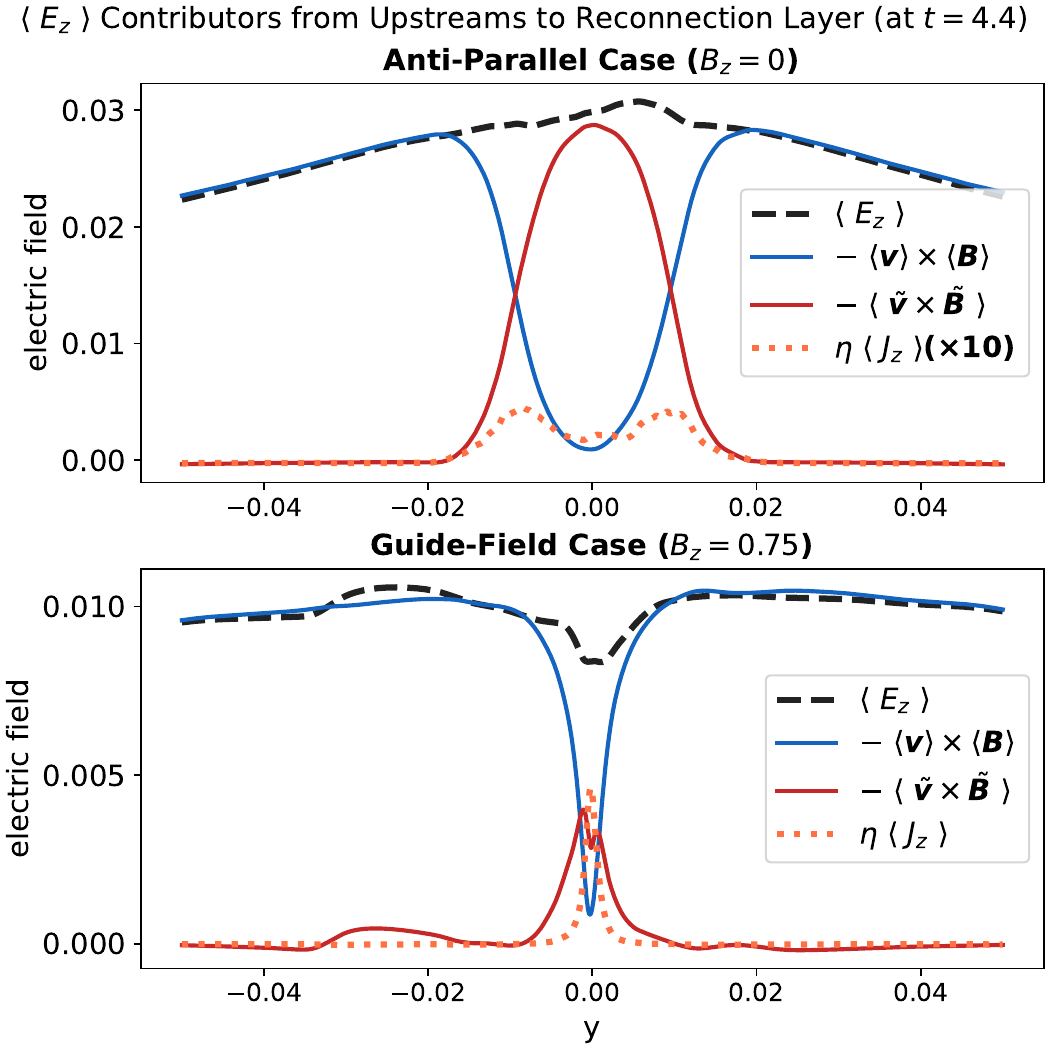}
    \caption{Contributions to the mean reconnection electric field $\langle E_z \rangle$. 
    Quantities are averaged along $z$ for fluctuation analysis and then averaged over $-0.1\leq x \leq 0.1$ to maximize signal to noise, obtaining 1D profiles in $y$. 
    In both cases, the dominant contribution shifts from mean-flow induction in the inflow region to the fluctuation EMF and resistive electric field inside the layer. 
    In the anti-parallel case (top), the reconnection layer is broader and $\langle E_z \rangle$ is dominated by fluctuation EMF at a level of $\simeq 0.03$. 
    In the guide-field case (bottom), the layer is narrower and at $y=0$ the turbulent EMF and resistive term make comparable contributions to $\langle E_z \rangle \simeq 0.01$.}
    \label{fig:emf}
\end{figure}

Figure \ref{fig:emf} plots $\langle E_z \rangle$ and its contributions. In the guide-field case (Figure \ref{fig:emf} bottom), the terms balancing the reconnection inflow are confined to a relatively narrow layer. At the midplane, comparable contributions from the fluctuation EMF and the resistive term combine with similar amplitudes to give a total $\langle E_z \rangle \simeq 0.01$. This phenomenology is consistent with a flux-rope-mediated interpretation, in which the resistive contribution is concentrated in narrow current sheets, while dynamic motions of locally coherent flux ropes provide the fluctuation EMF over a region that extends modestly beyond these sheets, similar to the mean fields for a 2.5D plasmoid chain \citep{beg2023}.

In the new anti-parallel case (Figure \ref{fig:emf} top), the reconnection layer is about three times broader (blue curve). The $\langle E_z \rangle$ is sustained at a level of $\approx0.03$, almost entirely by the fluctuation EMF (the resistive contribution has been multiplied by ten for plotting). Meanwhile, the non-ideal resistive contribution is spread across this broad layer, enabling reconnection throughout it.
Thus, the higher reconnection rate in the strong-shear regime is associated with a broader layer in which fluctuations provide most of the averaged electric field needed to balance the reconnection inflow, and current density is distributed across the whole layer instead of being concentrated at its center.

\section{Summary and Discussion} \label{sec:summary}

This Letter has shown that self-generated turbulent reconnection has two distinct regimes, controlled by the magnetic shear across the reconnection layer. 
In the guide-field, weaker-shear case, the MHD reconnection rate is $V_{\rm rec}/V_A \sim 0.01$, and the layer contains locally coherent flux ropes and thin current layers, consistent with recent theory \citep{russell2025} and simulations \citep[e.g.][]{daughton2011,huang2016,beg2022}. 
In the novel anti-parallel, strong-shear case, the current sheet instead develops short-wavelength corrugations that redistribute the current density over a broader region. When this happens, the MHD reconnection rate increases to $V_{\rm rec}/V_A \sim 0.03$, and the mean reconnection electric field is sustained mainly by turbulent EMF (rather than by a combination of a narrow current sheet and flux rope motions).

The discovery of this new regime of self-generated reconnection sets a new record for the reconnection rate in an MHD simulation with no external turbulent forcing. It also establishes magnetic shear as a key control parameter that delineates qualitatively different parts of the reconnection phase diagram for 3D systems. Interestingly, while the guide-field, weak-shear regime continues to be well described by the flux-rope-mediated theory provided by \citet{russell2025}, the faster anti-parallel, strong-shear regime instead departs from this picture. Its broader turbulent layer appears qualitatively similar to the stochastic-reconnection picture of \citet{lazarian1999}, but a quantitative comparison with LV99 is left for future work.

Another interesting observation is that once the initial translational symmetry breaks, the $v_z$ component of the velocity field develops significant fluctuations (up to the order $0.1$ in simulation units). Gradients of $v_z$ with respect to $y$ can drive MHD Kelvin-Helmholtz instabilities \citep[KHI;][]{chandrasekhar1961, kowal2020} that: grow fastest for short wavelengths (large $k$); 
produce dominant modes with $k_x \ll k_z$ in the zero-guide-field case; and are stabilized by sufficiently large $B_z$. These properties correspond well with the spatial structures seen in Figure~\ref{fig:j} (top), existence of distinct reconnection regimes controlled by the guide field, and occurrence of stronger fluctuation EMF for weak guide fields. These correspondences suggest that KHI may explain why and how MHD self-generated turbulent magnetic reconnection enters the newly-discovered faster regime for low guide fields, i.e. strong magnetic shear.

Finally, we highlight the importance of these findings for the triggering of fast reconnection in solar flares. Eruptive flare simulations show that the guide field can weaken by more than an order of magnitude during the eruption and vary strongly along the current sheet \citep{dahlin2022}. Thus, the shear-controlled transition to a faster reconnection regime may explain the sudden switch-on of fast reconnection during the eruption. This offers a potential solution to the onset of the flare impulsive phase and the rapid acceleration of CMEs.

%%%%%%%%%%%%%%%%%%%%%%%%%%%%%%%%%%%%%%%%%%%%%%%%%%%%%%%%%%%%%%%%%%%%%%%%%%%%%%%%

\begin{acknowledgments}
Q.M. is supported by a PhD scholarship awarded by the School of Mathematics \& Statistics at the University of St Andrews. A.J.B.R. acknowledges support from STFC grant ST/W001195/1. Simulations were run using the University of St Andrews HPC facility Hypatia.
The research made use of the Astrophysics Data System, which is operated by the Smithsonian Astrophysical Observatory under a NASA Cooperative Agreement, 80NSSC21M0056.
The authors thank P.F. Chen and the Nanjing University Solar Group, and R. Liu at the University of Science and Technology of China Solar Group, for hosting Q.M. to present this work in January 2026.
We thank the anonymous reviewer for constructive comments that improved the paper.
\end{acknowledgments}

%%%%%%%%%%%%%%%%

\begin{contribution}
% https://credit.niso.org
Q.M. ran the 3D simulations, performed the analysis and visualization, created the figures and drafted the manuscript. A.J.B.R. was responsible for the conceptualization of the study, design of its methodology, supervision, and reviewing the manuscript. Both authors edited the manuscript together.
\end{contribution}

%%%%%%%%%%%%%%%%

\facilities{University of St Andrews Hypatia High Performance Computing Centre}

\software{Lare3d \citep{arber2001},
          PyVista \citep{sullivan2019}}

\newpage

\bibliography{references}{}
\bibliographystyle{aasjournalv7}

\end{document}